\documentclass[aps,prl,twocolumn,groupedaddress,showpacs]{revtex4}

\usepackage{graphicx}

\newcommand{\rtHz}{\ensuremath\sqrt{\mbox{Hz}}}

\newcommand{\Vhz}{\mathrm{V}/\sqrt{\mbox{Hz}}}

\newcommand{\echz}{\ensuremath\mathrm{elementary\ charges}/\sqrt{\mbox{Hz}}}

\begin{document}

\author{S. E. Pollack\footnote{Present address: 
    Department of Physics and Astronomy and Rice Quantum Institute, 
    Rice University, Houston, TX  77251}
}
\author{M. D. Turner}
\author{S. Schlamminger}
\author{C. A. Hagedorn}
\author{J. H. Gundlach}
\affiliation{
    Center for Experimental Nuclear Physics and Astrophysics, University of Washington, Seattle, WA  98195}

\date{\today}

\title{Charge Management for Gravitational Wave Observatories using UV LEDs}

\begin{abstract} 
Accumulation of electrical charge on the end mirrors
of gravitational wave observatories, such as the space-based LISA
mission and ground-based LIGO detectors, can become a source of noise
limiting the sensitivity of such detectors through electronic couplings
to nearby surfaces.  Torsion balances provide an ideal means for
testing gravitational wave technologies due to their high sensitivity
to small forces.  Our torsion pendulum apparatus consists of a movable
Au-coated Cu plate brought near a Au-coated Si plate pendulum suspended
from a non-conducting quartz fiber.  A UV LED located near the pendulum
photoejects electrons from the surface, and a UV LED driven electron gun
directs photoelectrons towards the pendulum surface.  We have demonstrated
both charging and discharging of the pendulum with equivalent charging
rates of $\sim$$10^5\,e/\mathrm{s}$, as well as
spectral measurements of the pendulum charge resulting in a white noise
level equivalent to $3\times10^5\,e/\sqrt{\mbox{Hz}}$.  
\end{abstract}


\maketitle

%%%%%%%%%%%%%%%%%%%%%%%%%%%%%%%%%%%%%%%%%%%%%%%%%%%%%%%%%%%%%%%%%%%%%%%%%%%%%%%
%\section{Introduction} 
The freely floating proof masses in the Laser
Interferometer Space Antenna (LISA) will slowly acquire charge due
to cosmic ray impacts and solar particle collisions \cite{Stebbins}.
The accumulated charge will introduce forces on the proof mass by coupling
to residual DC electric fields from nearby surfaces as well as coupling
to the interplanetary magnetic field.  Fluctuations in these fields, in
addition to fluctuations in the charging rate, will provide a source
of force noise on the LISA proof mass that could affect the sensitivity of
gravitational wave observation.  Below $0.1\,$mHz this random charging
is expected to be the dominant source of acceleration noise \cite{merk}.
The baseline design to mitigate the proof mass charge is to shine UV light
on the proof mass and possibly the electrode housing surfaces.  The total
charge on the proof mass may then be reduced to acceptable levels through
the photoelectric effect from the incident UV radiation~\cite{merk}.

%%%%%%%%%%%%%%%%%%%%%%%%%%%%%%%%%%%%%%%%%%%%%%%%%%%%%%%%%%%%%%%%%%%%%%%%%%%%%%%
\begin{figure}[b!]
  \includegraphics[width=.9\columnwidth,angle=0]{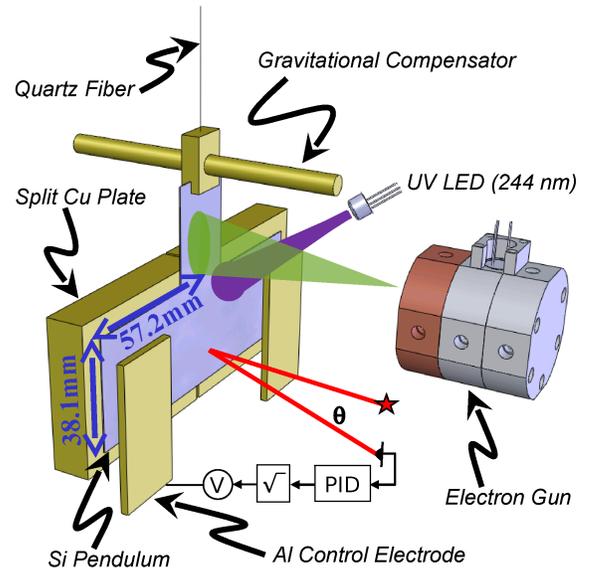}
\caption{ (color online)
Schematic showing the geometry of the apparatus.  A Si pendulum
is suspended from a quartz fiber with split Cu plate and Al electrodes
nearby.  The torsional displacement of the pendulum is measured optically
and is used in a feedback loop to control the pendulum angle.  Also shown
are the UV LED and the electron gun, which are used to 
remove and add electrons to the pendulum, respectively.  
\label{fig:setup}} 
\end{figure}
%%%%%%%%%%%%%%%%%%%%%%%%%%%%%%%%%%%%%%%%%%%%%%%%%%%%%%%%%%%%%%%%%%%%%%%%%%%%%%%

%%%%%%%%%%%%%%%%%%%%%%%%%%%%%%%%%%%%%%%%%%%%%%%%%%%%%%%%%%%%%%%%%%%%%%%%%%%%%%%
%%% new paragraph on LIGO
%%%%%%%%%%%%%%%%%%%%%%%%%%%%%%%%%%%%%%%%%%%%%%%%%%%%%

There exist similar concerns regarding electrical charging of the end
mirrors in the Laser Interferometer Gravitational-Wave Observatory (LIGO) \cite{LIGO}.  
The non-conducting fused silica mirrors in LIGO
are suspended from metallic frames, which interact electrically
with charge on the mirrors.  In Advanced LIGO, the mirrors will be
suspended in close proximity to reaction masses used to help maintain
a dark fringe in the interferometer \cite{AdvLIGO}.  Surface charge
fluctuations are therefore an important source of noise in the LIGO
interferometers and may significantly enter into the noise budget at low
frequencies \cite{LIGO_Noise, ICRC}.  Recent studies have shown that UV
illumination can be used to mitigate charge buildup on fused silica test
masses similar to those planned for Advanced LIGO \cite{LIGO_charge}.
In recent UV illumination tests, the discharging rate was measured to
be $<0.13\,$pC/(cm$^2$ s) and was dependent on the total charge on the sample,
resulting in non-linearities in the charging and a 
time constant of $\sim$1\,day \cite{LIGO_charge}.  
Given the high frequency band of LIGO
(10\,Hz---10\,kHz) it will be desirable to continue acquiring science data
during discharge events.  However, the viability of this is dependent
on the noise introduced from the discharging process.  In addition, the
UV radiation may alter the optical properties of the mirror coatings \cite{SunCoating, coating}.  
Investigations characterizing the noise
effects of discharging, such as the ones presented here, along with
effects on optical coatings, are warranted.

%%%%%%%%%%%%%%%%%%%%%%%%%%%%%%%%%%%%%%%%%%%%%%%%%%%%%%%%%%%%%%%%%%%%%%%%%%%%%%%

Our torsion-balance apparatus has been specifically designed to evaluate
forces between closely spaced conducting surfaces such as the LISA proof
masses and their housings, or the LIGO mirrors and reaction masses.
A schematic of our setup is shown in Fig.~\ref{fig:setup}.  It consists
of a rectangular Au-coated Si plate pendulum suspended from a thin fiber,
a movable Au-coated Cu plate of slightly larger size, and two Au-coated Al
control electrodes.  The torsional angle is measured by an autocollimator
that feeds back via a proportional-integral-derivative (PID) loop to
the voltage on one of the electrodes to keep the pendulum angle fixed.

\begin{figure}
  \includegraphics[width=0.75\columnwidth,angle=90]{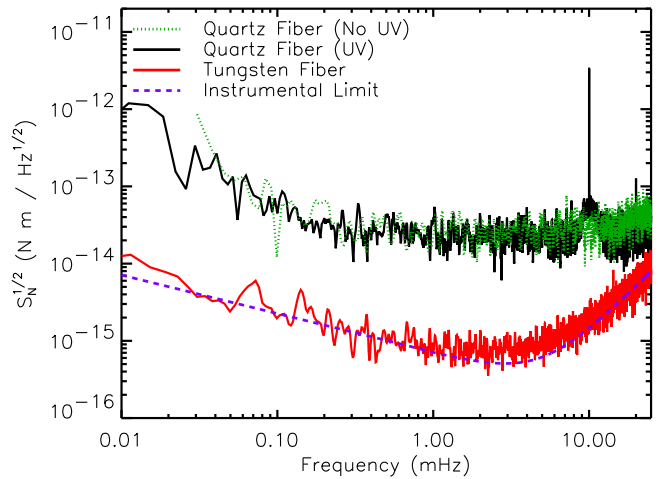}
\caption{ 
(color online) Torque sensitivity in feedback for a
plate-pendulum separation of 1\,mm with a tungsten (red)
fiber, quartz fiber (dotted green), and quartz fiber under UV
illumination (black).  The tones in the quartz spectrum are produced
from the charge measurement technique with a fundamental at 10\,mHz.
The instrumental limit when using the tungsten fiber (dashed) is a
combination of thermal noise (at low frequencies) and measurement noise
(at high frequencies).  The thermal noise of the quartz fiber is 1.7
times higher than that for the tungsten fiber due to the differences
in $\kappa$ and $Q$.  The measured noise level of the quartz fiber
is consistent with the measured charge fluctuations on the 
pendulum $\sim$2\,m$\Vhz$ \cite{surface}.  
\label{fig:noise}} 
\end{figure}

We have previously reported on the level of parasitic 
voltage fluctuations \cite{Stephan, surface} 
and thermal gradient related effects \cite{lisa6}
between closely spaced Au-coated surfaces.  In previous measurements,
the Si plate pendulum was suspended from a thin tungsten wire.
To electrically isolate the pendulum for charge measurements, the tungsten
wire has been replaced with a quartz fiber.  The quartz fiber was pulled
in an oxy-acetylene flame from a 1\,mm GE Type 214 fused quartz rod
using a mechanical device that rapidly pulled the spot-melted rod to
a fixed length of 53\,cm.  Since the fiber was pulled from a molten
liquid the diameter is tapered throughout the length of the fiber.
The mean diameter of the fiber is $\sim$$50\,\mu$m.  Torsional quality
factors of fibers such as this one have been measured to be on the
order of $10^5$ in the mHz regime \cite{Charlie}. 
The energy dissipation of the torsional motion is dominated by gas 
pressure damping, limiting the quality factor to $Q = 7.3(2)\times10^4$.
The torsional constant of the fiber is $\kappa = 41\,\mathrm{nN\,m/rad}$.
For comparison, the previously used tungsten fiber 
had $Q = 4000$ and $\kappa = 0.774\,\mathrm{nN\,m/rad}$,
resulting in a thermal torque spectral amplitude of $S_N^{1/2} (f) = \sqrt{2 k_B T \kappa / (\pi f Q)}$, about 1.7 times lower than that for the quartz fiber.
Measured torque noise traces of the pendulum suspended from
both tungsten and quartz fibers are shown in Fig.~\ref{fig:noise}.

%%%%%%%%%%%%%%%%%%%%%%%%%%%%%%%%%%%%%%%%%%%%%%%%%%%%%%%%%%%%%%%%%%%%%%%%%%%%%%%
%%%%%%%%%%%%%%%%%%%%%%%%%%%%%%%%%%%%%%%%%%%%%%%%%%%%%%%%%%%%%%%%%%%%%%%%%%%%%%%

The work function of Au is very nearly 5\,eV \cite{Anderson}, 
corresponding to a wavelength of 248\,nm.  To drive electrons off
the pendulum we utilize an ultraviolet LED with a nominal peak
wavelength of $244$\,nm and a maximum UV output power of 0.5\,mW.
Similar UV LEDs have been used in other charge management experiments
producing charging rates up to $5 \times 10^5\,e/\mbox{s}$~\cite{kxsun}.
In those experiments, AC charge management was demonstrated by 
applying a $\pm0.6\,$V potential to a nearby Au-coated surface and driving both the
potential and the UV LED at 10\,kHz.  The direction of the photoelectrons,
toward or away from the electrically isolated surface, was controlled by
the phase difference between the two signals.

In this report, we demonstrate DC charge control of an electrically
isolated pendulum by continuously or periodically driving on or off charges.
To charge the pendulum positively a UV LED is used, mounted above one of
the Al electrodes such that the light shines on the pendulum.  The LED is
controlled with a pulse-width modulated 20\,mA current source at 10\,Hz
with a typical duty cycle of $0.1\%$.  
%The charging rate is measured to be linear in both the UV current and the duty cycle.  
The duty cycle was chosen for a charging rate on the order of $10^5\,e/\mathrm{s}$, exceeding
the LISA requirement \cite{sumner}.

%%%%%%%%%%%%%%%%%%%%%%%%%%%%%%%%%%%%%%%%%%%%%%%%%%%%%%%%%%%%%%%%%%%%%%%%%%%%%%%
\begin{figure}
  \includegraphics[width=.8\columnwidth,angle=0]{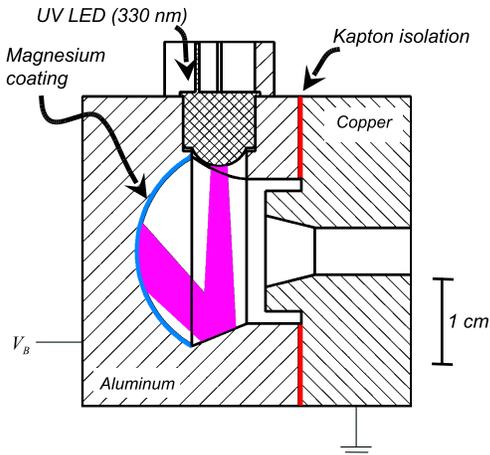}
\caption{ (color online)
Cross-sectional drawing of the electron gun.  UV light from the
LED creates photoemission from the magnesium coated cathode,
biased at $V_B = -6\,\mathrm{V}$.  Electrons accelerate
towards the grounded nozzle of the electron gun, maintain
their kinetic energy inside the field-free nozzle, and exit
the gun towards the pendulum.  
\label{fig:egun}} 
\end{figure}
%%%%%%%%%%%%%%%%%%%%%%%%%%%%%%%%%%%%%%%%%%%%%%%%%%%%%%%%%%%%%%%%%%%%%%%%%%%%%%%

Although electrons are efficiently removed from the pendulum using UV
light directed onto the surface, deposition of electrons by
illuminating nearby surfaces with UV light has not been successful.
To efficiently charge the pendulum negatively we use a low-energy electron gun.
The geometry of the gun~\cite{brewer} is shown in Fig.~\ref{fig:egun}.  
The gun consists of two
electrically isolated sections: a primarily Al cathode and a Cu nozzle.
The concave spherical reflector surface of the cathode is coated with Mg,
which has a work function of 3.66\,eV \cite{CRC}.  A UV LED with nominal
wavelength 330\,nm (3.76\,eV photon energy) is used to eject electrons
from the Mg surface.  With the nozzle grounded, electrons are focused
and accelerated by a negative bias voltage applied to the cathode,
usually $V_B = -6\,\mathrm{V}$.  
The geometry of the gun allows only a small amount of light
to escape through the nozzle, and the low photon energy
further prevents photoemission from the pendulum and nearby surfaces.
The LED in the electron gun is driven by a 20\,mA current source.

We model the plate-pendulum-electrode system as a collection of parallel-plate capacitors.
The charge $q$ on the pendulum can be determined from the voltage differences between
the pendulum and the nearby conducting surfaces forming the capacitors:
$q = \Sigma_i C_i (V_0 - V_i)$, 
where $V_i$ is the voltage on conducting surface $i$, 
$C_i$ is the capacitance of that surface with the pendulum,
and $V_0$ is the electric potential difference between the pendulum and instrumental ground.
The torque on the pendulum from each surface $i$ is given by
$N_i = \frac{1}{2}(dC_i/d\theta) (V_0-V_i)^2$,
and when operated in feedback mode $N_0 + \sum_i N_i = 0$, 
where $N_0 \approx 100\,\mathrm{pN\,m}$ is produced by the fiber.
In practice, three of the four nearby conducting surfaces
(both halves of the Cu plate and one control electrode)
are grounded, while the voltage of the remaining surface
is adjusted to control the angular position of the pendulum.

Since the torque is quadratic with the electric potential, 
two values of the control voltage will produce the same torque on the pendulum, 
labeled $V_+$ and $V_-$.
Because the system is in feedback, one finds that
for an uncharged pendulum $V_\pm$ are symmetric about zero,
whereas for a pendulum with charge $q$ they are symmetric about
$q/C$, where $C$ is the capacitance between the pendulum and half of the Cu plate.
The charge can then be determined by switching the
control voltage from one value to the other and calculating $q = C (V_+ + V_-)$.
This model has been verified with a finite element analysis using the geometry of the apparatus.
In practice, an iterative feedfoward scheme is used 
to determine a guess for the next switched value of the control voltage;
control of the pendulum is then maintained by the PID loop 
and $V_+$ or $V_-$ is measured before the next switching event.

The capacitance between the pendulum and half the Cu plate 
at a plate-pendulum separation of 4\,mm is $C=5.0(1)\,$pF.  The total capacitance of
the pendulum to ground is twice this amount plus a contribution
($<3$\,pF) due to the control electrodes.  
Surface potential differences $\delta V$
between the pendulum and the nearby conducting surfaces will modify
the above relation for $q$ by amounts proportional to $C \delta V$. 
Therefore our determination of charge noise $S_q^{1/2}$ is an upper limit for the
charge fluctuations on the electrically isolated pendulum.

\begin{figure}
  \includegraphics[height=1\columnwidth,angle=90]{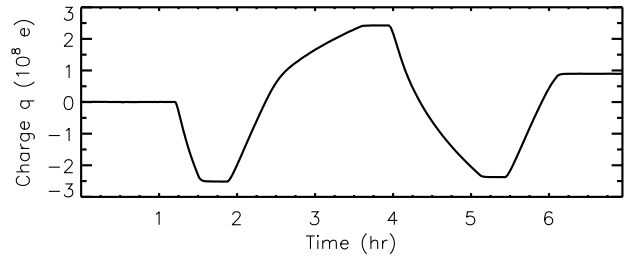}
\caption{ 
Charging and discharging the electrically isolated Si pendulum.
The charge on the pendulum is driven negatively (to $\,-2.5\times10^8\,e$) using
the electron gun and positively (to $\,+2.5\times10^8\,e$) using the UV LED.
\label{fig:charge}} 
\end{figure}

Figure~\ref{fig:charge} shows characteristic charging data.  The charge
rate is initially $2\times 10^5\,e/\mathrm{s}$ for charging both
positively and negatively, and the rate decreases as the charge on the
pendulum deviates from ground with a time constant of $\sim$40\,minutes.
GEANT4 simulations of energetic cosmic ray impacts with
the LISA spacecraft show a worst case charging rate of about $100\,e/\mathrm{s}$ \cite{Shaul} and
a more likely charging rate of about $50\,e/\mathrm{s}$ \cite{Shulte}, 
well within the charging capabilities demonstrated here. 

%%%%%%%%%%%%%%%%%%%%%%%%%%%%%%%%%%%%%%%%%%%%%%%%%%%%%%%%%%%%%%%%%%%%%%%%%%%%%%%
%%%%%%%%%%%%%%%%%%%%%%%%%%%%%%%%%%%%%%%%%%%%%%%%%%%%%%%%%%%%%%%%%%%%%%%%%%%%%%%

After neutralizing the charge on the pendulum to within 
$\pm 5 \times 10^6\,e$, we observe an ambient long term charging rate
of $\sim$$600\,e/\mathrm{day}$.  This charging rate was measured after
5 months in vacuum.  The initial charging rates after pumping down and
baking were on the order of $10^4\,e/\mathrm{day}$ and gradually decreased over
time.  We believe that this decrease in charging rate is indicative of a
redistribution of charges inside the vacuum chamber.  A similar charging
rate of $\sim$$3 \times 10^4\,e/\mathrm{day}$ has been observed in fused
silica optics \cite{Mitrofanov}, as well as a similar gradual decrease
in charging rate over time on fused silica samples \cite{Prokhorov}.
In addition to the measured slow change in charge on the pendulum, we
have observed dramatic changes in the charge, similar to those observed
in previous experiments \cite{Mitrofanov}, with some as large as $10^6$
elementary charges in 5 minutes.
We speculate that these events may be caused by high 
energy muons or cosmic rays \cite{Braginsky}; 
their actual source and temporal distribution require further investigation.

\begin{figure}
  \includegraphics[width=0.75\columnwidth,angle=90]{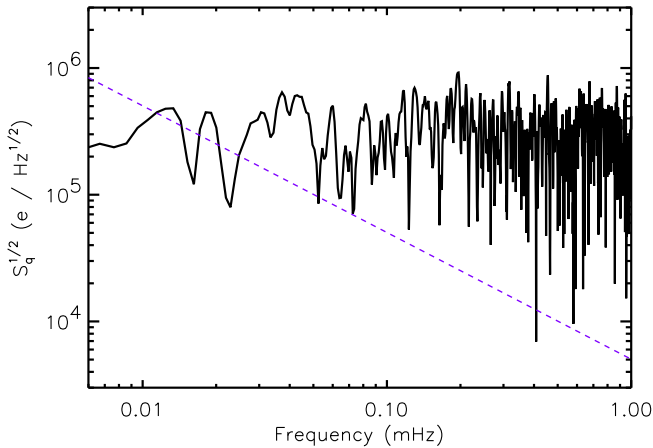}
\caption{ 
The measured charge noise (solid) in our experiment with an electrically isolated pendulum.  
The LISA requirement (dashed) is the contribution
of charge noise accounted for in the LISA acceleration budget,
assuming an effective charging rate of 500\,e/s.  
% The electrically
%measured Cu plate voltage noise (red), converted into an effective
%charge noise, represents one of the systematic noise sources in
%the measurement \cite{surface}.  
\label{fig:enoise}} 
\end{figure}

A spectrum of measured charge fluctuations on the pendulum after
neutralization is shown in Fig.~\ref{fig:enoise}.  The LISA requirement
in $e/\rtHz$, given by $S_q^{1/2} = \sqrt{2 \lambda} / (2 \pi f)$ with
$\lambda = 500 /$s the assumed effective charging rate, is the amount of
the acceleration noise budget currently allocated towards random charging
of the proof mass \cite{DRS_ITAT}.  We measure a white spectrum with a
voltage noise level $\sim$$2\,\mathrm{m}\Vhz$, equivalent to
$3\times10^5\,e/\rtHz$---meeting the LISA requirement at the lowest of frequencies.  
This measurement should be seen as an upper limit. 

It is possible that part of this noise may be caused by thermally activated
charge movement along the quartz fiber that would not occur in LISA.
Measuring the decay of the voltage on the pendulum, or the drift of
charge from an applied voltage on the support shaft of the quartz fiber,
allows us to calculate the resistance to ground of the pendulum through
the quartz fiber to be $1.1(3) \times 10^{19}\,\Omega$. 
A fundamental measurement limit is given by the Johnson noise generated in this resistor. 
The quartz fiber is in parallel with the plate-pendulum capacitor so that 
the real part of the impedance of the network is given by 
$\mbox{Re}\left[ Z (f) \right] =R/(1+4 \pi^2 f^2 R^2 C^2)$.
The corresponding voltage fluctuation is $S_V^{1/2} (f) = \sqrt{4 k_B T\, \mbox{Re}\left[Z (f)\right]}$~\cite{leach}.
Therefore the contribution of Johnson noise at 10\,$\mu$Hz is $\sim$60\,$\mu\Vhz$, about 30
times lower than the measured level $\sim$$2$\,m$\Vhz$.

%%%%%%%%%%%%%%%%%%%%%%%%%%%%%%%%%%%%%%%%%%%%%%%%%%%%%%%%%%%%%%%%%%%%%%%%%%%%%%%
%%%%%%%%%%%%%%%%%%%%%%%%%%%%%%%%%%%%%%%%%%%%%%%%%%%%%%%%%%%%%%%%%%%%%%%%%%%%%%%

Between midnight and 6\,AM, the root-mean-square level of the torque noise is a factor
of two smaller than during the daytime.  We have observed a similar modulation
in the torque noise of our pendulum when charge measurements are not undertaken.  
We did not observe this torque noise modulation when we used a tungsten fiber.
A separate torsion balance using a tungsten fiber, located in the vicinity
of this experiment, also observes a daily modulation of torque noise and therefore it
seems unlikely that this effect is due to a daily modulation in charging.
It is possible that this modulation is due to vibrational noise, which
was more efficiently damped when the lighter tungsten fiber was used.
We continue to investigate the source of this effect.  Effects such
as these clearly limit charge sensitivity and therefore the charge noise
measured represents an upper bound on the charge noise that could be
expected for Advanced LIGO and LISA.

We have demonstrated the feasibility of using UV LEDs for management
of charge in gravitational wave observatories such as LIGO and LISA.
Our particular setup implements DC charge control whereby we 
shine UV light or electrons on the pendulum to discharge it to near zero potential. 
One drawback of DC charge control is that discharging events may
introduce forces that result in complications for data acquisition or analysis \cite{gaps}.
However, the measurements presented here show no increase in the 
torque noise of the pendulum when discharging.
In contrast with AC charge control~\cite{kxsun,Wass,thesis},
DC charge control allows all surfaces surrounding the
target to be grounded, 
thereby minimizing the effects of stray voltage fluctuations on the pendulum~\cite{surface}.  
These measurements have been carried out with a torsion
pendulum suspended from a quartz fiber. 
The power spectral amplitude of the torque acting on the 
pendulum is a factor of ten higher than that expected
from the thermal noise model of a harmonic oscillator. 
Interestingly, this excess noise is consistent with fluctuation of the total charge on
the pendulum, which is measured to be a fraction of a million $\echz$, 
just above the current LISA charge noise requirement.  
Given the white nature of the measured noise and
the observed daily modulation, it is likely that this noise is not all
due to cosmic ray charging of the pendulum and therefore represents an
upper limit to the level that can be expected for gravitational wave observatories.

%%%%%%%%%%%%%%%%%%%%%%%%%%%%%%%%%%%%%%%%%%%%%%%%%%%%%%%%%%%%%%%%%%%%%%%%%%%%%%%

\begin{acknowledgments} 
We thank the members of E\"{o}t-Wash and the Center for Experimental Nuclear Physics
and Astrophysics at the University of Washington for infrastructure.
This work has been performed under contract NAS5-03075 through GSFC and
NASA Beyond Einstein grant NNG05GF74G.  
\end{acknowledgments}

\bibliography{charge}

\end{document}